\documentclass[twocolumn,showpacs,preprintnumbers,amsmath,amssymb]{revtex4}

\usepackage{graphicx}
\usepackage{dcolumn}
\usepackage{bm}
\usepackage{amssymb}

\begin{document}


\title{The Dynamic Transition of Protein Hydration Water}

\author{W.~Doster, S.~Busch and A.~M.\ Gaspar}
 \affiliation{Physik Department E 13 and ZWE FRM II, 
Technische Universit\"at M\"unchen, 85747 Garching, Germany}
 \email{wdoster@ph.tum.de}
\author{M.--S.\ Appavou, and J.~Wuttke}
 \affiliation{Forschungszentrum J\"ulich, JCNS at FRM II,
Lichtenbergstr.~1, 85747 Garching, Germany}
\author{H.~Scheer}
 \affiliation{Botanisches Institut II, Ludwig-Maximilians-Universit\"at
M\"unchen, Menzingerstr.~67, 80638 M\"unchen, Germany}

\date{\today}

\begin{abstract}
Thin layers of water on biomolecular and other nanostructured surfaces
can be supercooled to temperatures not accessible with bulk water.
Chen \textit{et al.} [PNAS 103, 9012 (2006)] suggested that anomalies near 220~K observed by quasi-elastic neutron scattering can be explained  by a hidden critical point of bulk water. Based on more sensitive measurements of water on perdeuterated phycocyanin, using the new neutron backscattering spectrometer SPHERES, and an improved data analysis, we present results that show no sign of such a fragile-to-strong transition. The inflection of the elastic intensity at 220~K has a dynamic origin that is compatible with a  calorimetric glass transition at 170~K. The temperature dependence of the relaxation times is highly sensitive to data evaluation; it can be brought into perfect agreement with the results of other techniques, without any anomaly.
\end{abstract}


\maketitle

In contrast to bulk water,
protein hydration water
can be supercooled down to a glass transition at $T_{\rm g}\simeq 170$~K.
Near $T_{\rm g}$
translational degrees of freedom arrest,
which induces discontinuities in the specific heat
and the thermal expansion coefficient of the hydration water
\cite{Doster10,Miyazaki00,Demmel97,Jansson10}.
Due to the dynamic nature of the glass transition,
freezing of microscopic degrees of freedom can already be observed
far above $T_{\rm g}$.

The \textit{protein dynamic transition} is an abrupt onset of
atomic displacements on the microscopic length and time scale
probed by quasielastic neutron scattering (QENS).
First observed twenty years ago in hydrated myoglobin and lysozyme 
at $T_\Delta \simeq240$~K \cite{Doster89},
it is now known to be a generic property of hydrated proteins,
while it is absent in dehydrated systems. 
It is therefore related to the dynamics of the hydration shell.

In QENS, mean squared dispacements $\left\langle \delta x^2 \right\rangle$
are deduced from the elastic scattering intensity $S(q,0)$.
Due to the finite spectrometer resolution (fwhm $2\Delta\omega$),
one actually measures $S(q,|\omega|\lesssim\Delta\omega)$.
Full spectral measurements show that the anomalous decrease of the central
peak is compensated by increasing inelastic wings
\cite{Doster89,Doster90}.
These effects have been interpreted
as precursors of the glass transition.
Since QENS probes structural relaxation at $\Delta\omega^{-1}\simeq100~$ps,
while the calorimetric $T_{\rm g}$ refers to a time scale of 100~s,
it is natural that $T_\Delta$ is located far above $T_{\rm g}$:
the protein dynamic transition
is the microscopic manifestation of the glass transition
in the hydration shell.
The time-scale dependence of $T_\Delta$ also explains its
variation with viscosity \cite{Lichten99,Doster05,Doster08}.

Recently, these views have been challenged by the suggestion
that there might be a time-scale independent transition.
Support came mainly from QENS experiments
on the backscattering spectrometer HFBS at NIST.
In hydrated lysozyme, DNA and RNA,
a kink was found not only in $S(q,|\omega|\lesssim\Delta\omega)$,
but also in the $\alpha$ relaxation time $\left\langle \tau \right\rangle$
deduced from full spectra $S(q,\omega)$
\cite{Chen06,Chen06a,Chu08,Chu09}.
This change of $\left\langle \tau \right\rangle(T)$
from high-$T$ super-Arrhenius 
to low-$T$ Arrhenius behavior 
at $T_{\rm L}\simeq220$~K has been interpreted as
a \textit{fragile-to-strong-transition (FST)}
from the high density (HDL) to the less fluid low density phase (LDL)
of supercooled water \cite{Poole92}.
In this view, a qualitative change in the dynamics 
occurs when the so-called Widom line is crossed,
which extrapolates the phase boundary beyond the conjectured critical point
\cite{Ito99,Xu05}.

Similar behaviour was found
for confined water
in various non-biological environments
[\onlinecite{Mamontov08b}, \onlinecite{Mamontov09} and references therein];
$T_{\rm L}$ was mostly located between 215 and 228~K.
There were, however, two remarkable exceptions:
making a substrate hydrophobic lowered $T_{\rm L}$ by 35~K \cite{Chu07};
using another neutron spectrometer with wider dynamic range
lowered $T_{\rm L}$ by about 10 to 20~K \cite{Mamontov08a}.
Critics objected that relaxation times measured
by other spectrocopic techniques
show no anomaly at $T_{\rm L}$ \cite{Swenson06,Vogel08,Pawlus08}.

In this paper we will argue that the 
changing temperature dependence of
log($\left\langle \tau \right\rangle$)
is likely to be an artifact of the data analysis,
and that the kink in the elastic intensity
also admits a conventional, \textit{dynamic} interpretation.
We will refer to the hydration shell of different proteins.
If the anomalies observed around $T_{\rm L}$ were due to
an intrinsic property of water,
then the choice of the protein should not matter.

\begin{figure}
\centerline{\includegraphics*[width=3.4in]{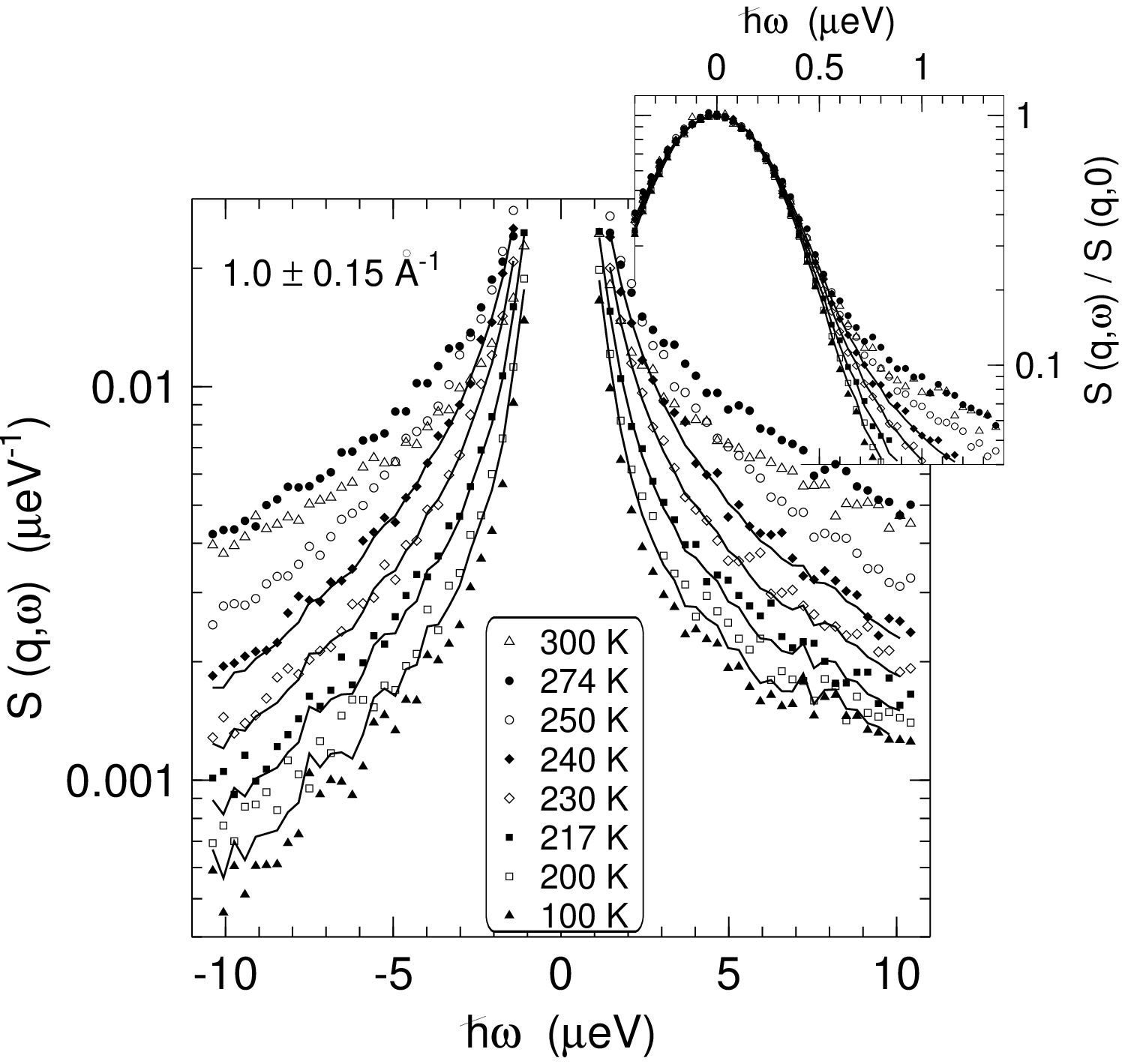}}
\caption{Neutron scattering spectra of H$_2$O on d-CPC at
$q=1.0\pm0.15$~\AA$^{-1}$. Note the truncated logarithmic scale.
Solid lines are fits with~(1),
convoluted with the instrumental resolution
(sample data at 100~K). ---
Inset: central peak, normalized to the value at $\omega=0$.}
\label{fig_1}
\end{figure}

To improve the empirical base,
we measured the motion of protein hydration water with enhanced sensitivity,
using a new neutron spectrometer and a fully deuterated protein.
We chose C-phycocyanin (CPC),
a light harvesting, blue copper protein.
Deuteration drastically reduces scattering from the protein
so that we become more sensitive to weak quasielastic scattering
from the hydration water.
CPC has been investigated by neutron scattering
before \cite{Middendorf80,Bellissent92,Koeper07,Gaspar09,Bradley94}.
The sample was purified from a preparation 
of Crespi \cite{Bradley94,Crespi} with nearly 99 \% deuteration.
The absorbance ratio
$A_\text{620\,nm}/A_\text{280\,nm}\simeq7.3$
is above the analytic-grade value of~5 \cite{Patil06}.
The integrity of the secondary structure was ascertained
by CD spectroscopy. The protein was dialyzed against distilled water, freeze-dried and rehydrated
to h = 0.3~g/g \cite{Gaspar09}, the same degree of hydatrion as in Ref.~\cite{Chen06}.

QENS was performed on 500 mg of hydrated D-CPC
with the new backscattering spectrometer SPHERES at FRM~II \cite{Wuttke09}.
It has a resolution (fwhm) of 0.62~$\mu$eV (HFBS 0.85~$\mu$eV),
high flux, and an outstanding signal-to-noise ratio of 1000:1.
Scattering wavenumbers $q$ range from 0.2 to 1.8~\AA$^{-1}$.
We chose about the same energy range $\pm10$~$\mu$eV
as customary at HFBS.
Spectra at temperatures $T$ between 100 and 300~K were
collected for 6 to 12~h.
An empty cell measurement was subtracted
and spectra were normalized to a vanadium standard.
The analysis was performed with the program \textsc{Frida} \cite{Frida}.

Fig.~\ref{fig_1} displays selected spectra
in the relevant temperature range.
The logarithmic intensity scale emphasizes
both the broad, asymmetric wing of instrumental resolution function,
unavoidable in a crystal spectrometer,
and the quasielastic scattering.
At least two distinct components of the scattering function $S(q,\omega)$ can be observed:
(i) An elastic component that
has the shape of the instrumental resolution function $R_q(\omega)$.
It is mainly due to the coherent structure factor of the protein
(as evidenced by strong small-angle scattering for $q\ll1$~\AA$^{-1}$ \cite{Gaspar09}),
as well as to a protein incoherent term mainly due to the
exchanged protons (20\%).
Also immobilized water near charged groups
contributes to this component.
(ii) A quasielastic component that appears above about 170~K.
It is mainly due to motion of adsorbed H$_2$O molecules.

\begin{figure}
\includegraphics*[width=2.3in]{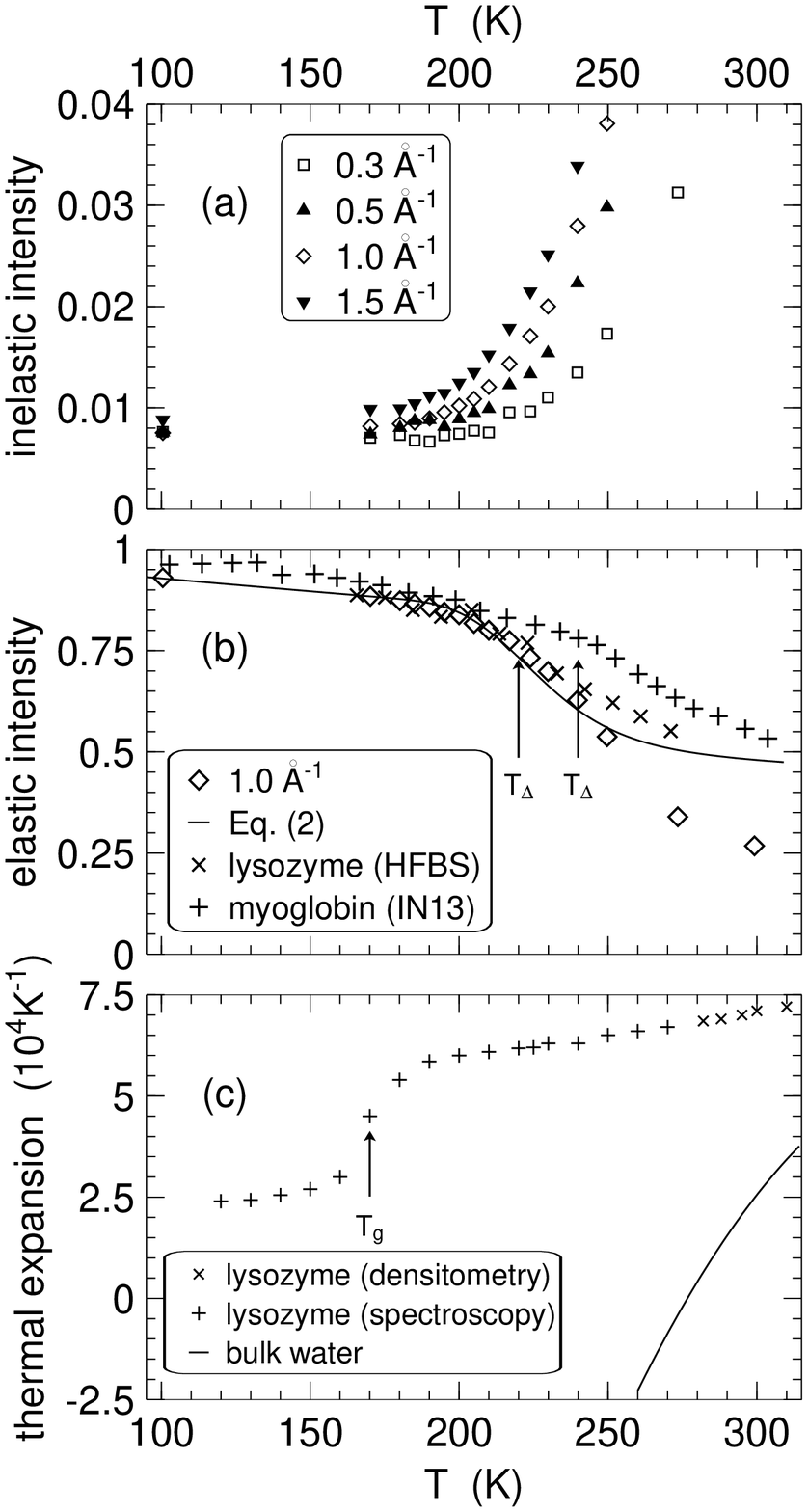}
\caption{(a) Scattering intensity integrated over
$\hbar\omega<-3$~$\mu$eV.
(b) Elastic scattering intensity $S(q,0)$ and result of~(2),
compared to literature data \cite{Chen06,Doster89}.
(c) Thermal expansion coefficient $\alpha_{\rm P}$
of water in hydrated lysozyme powder
($+$, spectroscopic measurement \cite{Demmel97}),
of hydration water in lysozyme solution
($\times$, densitometry, \cite{Hiebl91}),
of bulk water (line).}
\label{figC}
\end{figure}

To start with a model-free data characterization,
we examined the $T$ dependence of the integrated inelastic intensity
for different $\omega$ intervals,
preferentially on the $\omega<0$ side where the resolution wing is weaker.
Between 180 and 230~K, the inelastic intensity begins to rise strongly.
There is no sharp kink;
the temperature of maximum inflection depends
on the $\omega$ range and on~$q$ (Fig.~2a).
This is a first indication for a dynamic phenomenon rather
than a hidden phase transition:
the $q$ and $T$ dependence is compatible
with a smooth onset of translational diffusion.
The elastic intensity in compensation shows a step-like decrease at 220~K in parallel with data obtained with lysozyme. This suggests a common transition temperature of both hydrated proteins (Fig.~2b).
In the following,
we concentrate on $q=1.0\pm0.15$~\AA$^{-1}$.

To analyse the spectra, we use the same model as Chen \textit{et al.}\ \cite{Chen06}, consisting of an elastic line and a 
Kohlrausch-Williams-Watts function
(KWW, Fourier transform of a stretched exponential \cite{libkww}):
\begin{equation}
S_{\rm th}(q,\omega) =  f_q\left\{ a_1\delta(\omega) + 
          a_2\int_{-\infty}^\infty\!\frac{{\rm d}t}{2\pi}\,
                   {\rm e}^{i\omega t-(|t|/\tau)^{\beta}}\right\}.
\end{equation}
The Debye-Waller factor $f_q=\exp(-\left\langle \delta x^2 \right\rangle q^2)$
accounts for phonon scattering outside our energy window.
The mean squared displacement $\left\langle \delta x^2 \right\rangle$ is assumed
to be linear in $T$;
from 100 and 170~K:
$\partial\langle \delta x^2\rangle/\partial T=7\cdot10^{-4}$~\AA$^2$/K.
$\beta$ was fixed to 0.5 as in the analysis of Chen \textit{et al.}

Equation (1) can fit the data sets for $T \leqslant 240$~K, hereby covering the temperature range of the proposed transition.
At higher $T$, the data start to deviate from Eq.~(1):
additional quasielastic scattering, likely due to relaxational or fast anharmonic motion of the protein, is not included in this bimodal model.

In a first attempt to extract the relaxation times, we have used a standard fitting procedure for the data evaluation.
Fitting the quasielastic wings for $T<240$~K,
we are in the regime $\omega\tau\gg1$
where the KWW function
goes in first order with $(\omega\tau)^{-1-\beta}$.
In such a power law regime,
the fit parameters amplitude and time scale $\tau$ are degenerate.
Therefore, the 240~K data were used to determine the relative contribution of the water $a_2/(a_1+a_2)$ to 0.42 which was then used as input for the other temperatures.


However, this approach neglects a severe limitation of the standard fitting procedure for narrow lines:
When fitting the central peak,
the convolution of $S_{\rm th}$ with the experimental resolution $R_q(\omega)$
becomes nontrivial for $T\lesssim220$~K
where the peak in the KWW function becomes so sharp
(compared to the experimental energy binning)
that the straightforward calculation $S_{\rm th}(q,\omega)\otimes R_q(\omega)$
as a Riemann sum yields a wrong amplitude $a_1$.
Because of the mentioned degeneracy of amplitude and time scale, this results in a distorted $\left\langle \tau \right\rangle$.
We have therefore repeated the fits with an improved fitting procedure, neglecting the erroneous $a_1$ and fixing directly $a_2$ at 0.42.
\begin{figure}
\centerline{\includegraphics*[width=3.in]{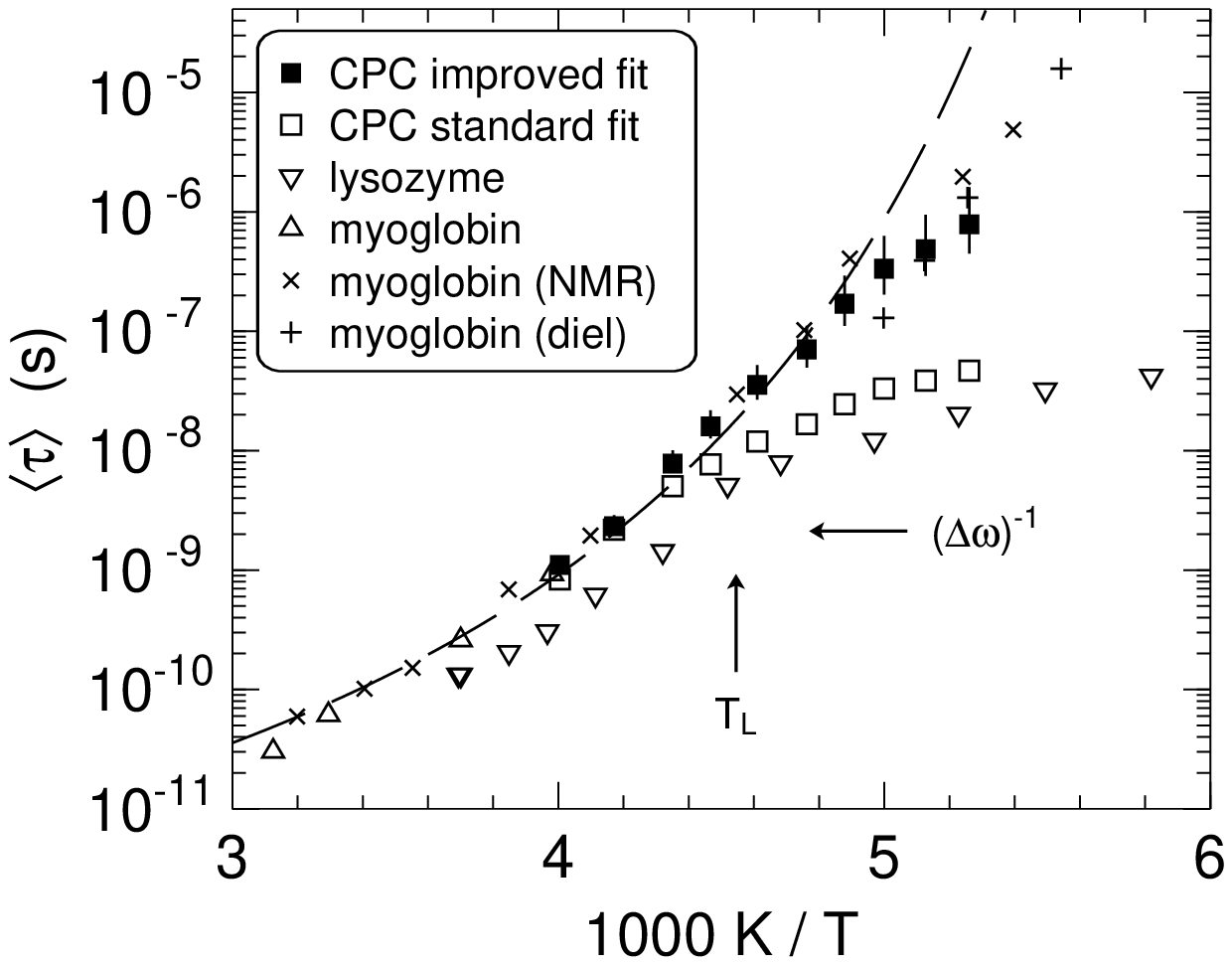}}
\caption{Average relaxation times $\left\langle \tau \right\rangle$
versus reciprocal temperature for hydration water on different proteins:
($\blacksquare$): c-phycocyanin with improved fitting procedure and
($\square$): with a standard fitting procedure. Error bars were estimated by varying $\beta$ between 0.45 and 0.55.
($\bigtriangledown$): lysozyme, QENS \cite{Chen06}.
($\bigtriangleup$): myoglobin, QENS with wide $\omega$~range \cite{Doster05}.
($\times$): myoglobin, NMR \cite{Lusceac10}.
($+$): myoglobin, dielectric loss \cite{Jansson10},
vertical arrow: conjectured $T_{\rm L}$ of the FST,
horizontal arrow: instrumental resolution $(\Delta \omega)^{-1}$.
Dashed line: Vogel-Fulcher fit to the data above 200~K with fixed $T_{\rm g}=170$~K, yielding $\left\langle \tau \right\rangle_0 = 6.5 \cdot 10^{-13}$~s, $T_{\rm H} = 745$~K, $T_0 = 147$~K.}
\label{figB}
\end{figure}


Fig.~3 shows an Arrhenius plot of the mean relaxation time
$\left\langle \tau\right\rangle
 = \int{\rm d}t  \exp(-t/\tau)^{\beta}
 = \beta^{-1}\Gamma\left(\beta^{-1}\right)\tau$
resulting from both fitting procedures.
As can be seen, the temperature dependence of the relaxation times obtained from the standard fitting procedure resembles the results of Chen \textit{et al.}
This trend is however not observed any more in the results of the improved fitting procedure which agree well with
the structural ($\alpha$-)relaxation times of water on other proteins
taken from literature (also shown). 
There is a good agreement with high-$T$ myoglobin data measured by QENS over a wider
energy range than covered here~\cite{Doster05} and
for $T\lesssim220$~K with 
both, $^2$NMR \cite{Lusceac10}
and dielectric spectroscopy \cite{Jansson10} which show a smooth increase without inflection at 220~K.
The curvature of $\left\langle \tau \right\rangle$ in the Arrhenius plot indicates fragile behaviour according to a Vogel-Fulcher relation, $\left\langle \tau \right\rangle = \left\langle \tau \right\rangle_0 \exp \left[ T_{\rm H}/(T-T_0) \right]$.

For low $T$, our $\left\langle \tau\right\rangle$ exceeds the resolution $\Delta\omega^{-1}\simeq2$~ns by some orders of magnitude.
Still, for $|\hbar\omega|\gtrsim2$~$\mu$eV there are clear quasielastic wings
that allow to determine $\tau$ within the chosen model
with fixed parameters $a_2$ and $\beta$.
Therefore, this analysis can be used to probe the spectral behaviour for the predicted FST discontinuity.

Whereas our first analysis, using the standard fitting procedure, reproduces the results from hydrated lysozyme \cite{Chen06} -- the basis for the postulated FST at 220~K --
we can rule out any discontinuity at this temperature using the improved fitting procedure.
%
%
%
We suspect therefore that also the various results backing the FST are affected by the problems of the standard fitting procedure:
In the power-law regime $\omega\tau\gg1$,
fitted $\left\langle \tau \right\rangle$ are meaningful only if
the amplitude $a_2$ has been fixed at a physically reasonable value.
In the FST reports, the amplitude, if mentioned at all,
is just called a \textit{scaling constant} \cite{Mamontov08b}.
The fitted $\left\langle \tau \right\rangle(T)$
are therefore possibly distorted by an uncontrolled drift of $a_2(T)$.
This could explain why the cross-over temperature $T_{\rm L}$ varies with pressure, while the cross-over time appears constant \cite{Chu09}.

Below approximately 200~K, the relaxation times extracted from our improved technique deviate slightly from the Vogel-Fulcher relation. Being orders of magnitude above the instrumental resolution, a detailed discussion of this effect seems inadequate. If it were real, one could attribute it to the onset of the glass transition which is known to be about 30~K broad \cite{Jansson10,Khodadadi10}. The observed process would then no longer be the isotropic $\alpha$~relaxation \cite{Lusceac10}.

In contrast to the kink in $\left\langle \tau \right\rangle$, the kink in the observed elastic scattering intensity
$S(q,|\omega|\lesssim\Delta\omega)$
is an undisputable experimental fact.
Fig.~2b shows that observations at SPHERES (CPC) and
HFBS (lysozyme \cite{Chen06}) almost coincide below 240~K.
The difference at higher $T$ can be explained by our use
of deuterated CPC, which reduces the relative contribution of the protein.

The expected elastic scattering is
\begin{equation}
  \begin{array}{l}
   S_{\rm exp}(q,0) = \displaystyle
                       \int\!{\rm d}\omega\, S_{\rm th}(q,\omega)\,R_q(-\omega)\\
 = \displaystyle
    f_q \left\{ a_1 R_q(0) + (1-a_1)
   \int_0^\infty\!\frac{{\rm d}t}{\pi}\,
               {\rm e}^{-(t/\tau)^\beta}\,\tilde{R}_q(t) \right\}
  \end{array}
\end{equation}
with $\tilde{R}_q(t)=\int\!{\rm d}\omega \cos(\omega t)R_q(\omega)$.
Indeed, the low-$T$ data are perfectly reproduced by Eq.~(2)
using the values from the Vogel-Fulcher relation for $\left\langle \tau \right\rangle$.
This confirms the long established explanation
of the protein dynamic transition as
a time-scale dependent glass transition:
When $\tau^{-1}$ exceeds the instrumental resolution $\Delta\omega$,
then the central part of the KWW spectrum starts to broaden
so that $S(q,|\omega|\lesssim\Delta\omega)$ falls below~$f_q$.

This is further corroborated by elastic scattering ($q=1.4$~\AA$^{-1}$)
from hydrated myoglobin
on the thermal neutron backscattering spectrometer IN13 \cite{Doster89}.
Its resolution width $2\hbar\Delta\omega=8$~$\mu$eV
is an order of magnitude larger than that of HFBS or SPHERES,
and the kink in $S(q,|\omega|\lesssim\Delta\omega)$ is observed 
accordingly at a higher temperature,
namely the often reported $T_\Delta\simeq240$~K.

Previous QENS experiments were motivated by the hope
that hydration layers would provide an opportunity
to study water in a deeply supercooled state that is not accessible in the bulk.
However, thin hydration layers seem to be 
qualitatively different from bulk water
\cite{Swenson06}.
This view is also supported by literature data of
the thermal expansion coefficient $\alpha_{\rm P}$ of lysozyme hydration water
(Fig.~2c).
In concentrated solution, $\alpha_{\rm P}$
is nearly $T$ independent and much larger than the bulk value \cite{Hiebl91}.
In hydrated powder, $\alpha_{\rm P}$
is estimated from the O-H stretching frequency of water,
which varies with the average H bond length.
A striking step is found near 170~K,
induced by the softening of the O-H--O hydrogen bond network
\cite{Demmel97}.
No anomaly is observed at $T_{\rm L}\simeq220$~K.

All observations agree, however,
with a glass transition of the hydration shell
and a dynamic onset temperature $T_\Delta$, which varies with the probe frequency (Fig.~2b).
The different values of $\alpha_{\rm P}$ for bulk and hydration water
could indicate
that the protein acts as a patch-breaker \cite{Stanley80},
suppressing critical fluctuations present in bulk water.

This work has been supported by Deutsche For\-schungs\-gemein\-schaft
through SFB 533.


\end{document}